# STUDY OF THE ANOMALOUS SKIN EFFECT OF NORMAL CONDUCTING FILM

Binping Xiao[†], M. Blaskiewicz, T. Xin

*Brookhaven National Laboratory, Upton, New York 11973-5000, USA*

For the radiofrequency (RF) applications of normal conducting film with large mean free path at high frequency and low temperature, the anomalous skin effect differs considerably from the normal skin effect with field decaying exponentially in the film. Starting from the relationship between the current and the electric field (E field) in the film, the amplitude of E field along the film depth is calculated, and is found to be non-monotonic. The surface impedance is found to have a minimum value at certain film thickness. We apply this calculation into a Cu coated S.S. beam pipe used in an accelerator to reduce the ohmic power loss to determine the minimum thickness that should be applied.

## I. Introduction

Coating Stainless Steel (S.S.) beam pipe of particle accelerators with Cu was previously considered in a variety of projects to reduce the resistive wall impedance and the ohmic power loss[1-5]. In the Relativistic Heavy Ion Collider (RHIC) at Brookhaven National Lab (BNL), the vacuum chamber in the cold arcs is made of 316LN S.S.. The high wall resistivity can result in unacceptable ohmic heating that is dissipated into the cryo system. This is an even larger concern in the upgrade project electron ion collider (BNL EIC) as the beam is more intense[4, 5]. In this application, the time and cost of in-situ coating is a concern, thus analysis of the minimum film thickness requirement is needed. This paper starts from the anomalous skin effect of bulk material in section II, and extends it to the film material in section III using a method similar to previous references [6, 7], in section IV Cu film without substrate is used to explain the anomalous behavior, Cu film on S.S. substrate for BNL EIC is calculated in section V to get the minimum film thickness, and conclusion is shown in section VI. Please note this paper aims on theoretical studies, detailed thin film fabrication technologies are not shown, neither did we take engineering details, i.e., beampipe curvature, film flatness, surface roughness, into consideration.

## II. Anomalous skin effect of bulk normal conductor

In this section we briefly introduce the anomalous skin effect of bulk normal conductor with diffusive reflection. It was previously studied by Reuter and Sondheimer[8, 9]. In their study, a semi-infinite metal was considered, with its surface in the *xy*-plane and the positive *z*-axis directed towards the interior of the metal, and $z=0$ the interface between vacuum and metal. The electric field $E(z)e^{i\omega t}$ is in the *x*-direction. The relationship between $E$ and the current $J$ is [9]:

$$E''(z) + k^2 E(z) = i\omega\mu_0 J(z) \quad (1)$$

Here derivative is over depth $d/dz$, and $k=\omega/c$. For the anomalous skin effect, $J(z)$ is not solely determined by the local electric field $E(z)$, instead, it is [9]:

$$J(z) = \int_0^\infty k_a(z-z_1)E(z_1)dz_1 \quad (2)$$

with $\quad k_a(z) = \dfrac{3}{4\rho\ell}\int_1^\infty e^{-|z|sa/\ell}(\dfrac{1}{s}-\dfrac{1}{s^3})ds \quad$ and

$a = 1+i\omega\ell/v_F$, with $\ell$ the mean free path and $v_F$ the Fermi velocity.

The surface impedance can be calculated from the $E$ field on the surface $z=0$ and its derivative over depth: $Z = -i\omega\mu_0 E_0 / E_0'$. For bulk metal, instead of direct calculating $E(z)$, one can do fourier transform to the above equations and get the surface impedance [9, 10]. With diffuse reflection of electrons of the metal surface it is:

$$Z = i\pi\omega\mu_0 \left\{\int_0^\infty \ln[1+K(p)/p^2]dp\right\}^{-1} \quad (3)$$

where

$$K(p) = \dfrac{3\omega\mu_0}{2\rho\ell p^3}\left\{-\dfrac{ipa}{\ell} + [p^2+(\dfrac{a}{\ell})^2]ArcTanh[\dfrac{\ell p}{-ia}]\right\}$$

is the fourier transform of $k_a(z)$.

This equation can also be achieved using the Mattis-Bardeen theory for normal conductor [11, 12], by applying [9]

$$I(\omega,z) = -\pi i\hbar\omega e^{-iz\omega/v_F}$$

into [10]

$$K(p) = \dfrac{-3\mu_0}{4\pi\hbar v_F \Lambda(0)}\int_0^\infty \int_{-1}^1 e^{ipzu} e^{-\dfrac{z}{\ell}}(1-u^2)$$
$$\times I(\omega,z)dudz$$

With $\Lambda(0) = \dfrac{m}{N(0)e^2}$, and $N(0) = \dfrac{mv_F}{e^2\rho l}$ the free electron density.

## III. Anomalous skin effect of normal conducting film

For a normal conductor with finite thickness $d$, one needs to calculate the electric field over depth $E(z)$ to get the surface impedance. In this case the expression of $J(z)$ changes to [3]:

$$J(z) = \int_0^d k_a(z-z_1)E(z_1)dz_1 \qquad (4)$$

We use the dimensionless variable $u = z/\ell$, apply equation (4) into (1) and get [3]:

$$d^2E(u)/du^2 + k^2\ell^2 E(u) = i\omega\mu_0\ell^3 \int_0^{d/\ell} k_a(u\ell - u_1\ell)E(u_1)du_1 \qquad (5)$$

First assume at $z=d$, the boundary condition is $E(z)=E_d$ and $E'(z)=E_d'$. Integrating the above equation twice with respect to $u$ gives:

$$E(u) + \int_u^{d/\ell} du_2[(u-u_2)(k^2\ell^2 E(u_2) + i\omega\mu_0\ell^3 \int_0^{d/\ell} k_a(u\ell - u_1\ell)E(u_1)du_1)] \qquad (6)$$
$$= E_d - (d-u\ell)E_d'$$

To solve the above equation numerically, the depth (normalized to $\ell$) is divided into $K$ equal segments $\Delta = d/\ell/K$. We use $E_n$ to note the $E$ field at $n\Delta$, thus we have:

$$E_n + \sum_{m=n}^K (n-m)[k^2\ell^2 E_m + i\omega\mu_0\ell^3\Delta^3 \sum_{j=0}^K E_j k_a((m-j)\ell)] \qquad (7)$$
$$= E_d - \ell\Delta(K-n)E_d'$$

$k_a(0)$ is infinite and is replaced by $\dfrac{1}{\Delta}\int_{-\Delta/2}^{\Delta/2} k_a(z)dz$.

The above equation can be solved by matrix inversion to get $E_n$, with the results normalized to $E_d$ ($E_d=1$). Similar to the bulk case, the surface impedance of this film is $Z = -i\omega\mu_0 E_0/E_0'$.

Here we briefly discuss the boundary conditions with different substrate.

1) with a bulk metal substrate with its surface impedance, noted as $Z_{sub}$, in the normal skin effect regime. The boundary condition is $E_d' = -i\omega\mu_0 E_d/Z_{sub}$.
2) without substrate. In this case $Z_{sub} = 377\Omega$.
3) with PEC boundary. In this case $E_d = 0$ and $E_d' \neq 0$.
4) with a bulk dielectric substrate. In this case the characteristic impedance of the substrate is used to the above boundary condition, with $Z_{sub} = Z_I = \dfrac{Z_{vac}}{\varepsilon_r^{1/2}}(1 + \dfrac{i}{2}\tan\delta)$, with $Z_{vac}=377\Omega$, $\varepsilon_r$ the relative electric permittivity and $\tan\delta$ the loss tangent, of the substrate.
5) with a dielectric substrate (characteristic impedance $Z_I$) from $z=d$ to $z=D$, and then a bulk metal with surface impedance $Z_{subD}$ after $z=D$. In this case the transmission line method [13] is used to get the substrate surface impedance: $Z_{sub} = \dfrac{Z_{subD} + iZ_I \tan(\beta(D-d))}{Z_I + iZ_{subD}\tan(\beta(D-d))}Z_I$, with $\beta = \omega\mu_0/Z_I$ the complex wave number of the plane wave penetrating into the dielectric. The method used in [13] to calculate the field in a metal is not applicable to our case since it is valid only in normal skin effect. To avoid the resonance in the substrate, the thickness of dielectric should be chosen so that $D-d = \dfrac{c}{2f\varepsilon_r^{1/2}}n$, $n=1,2,3\ldots$ does not happen [13].

In the following sections, the first two cases are discussed in detail. Cu is used as an example. The parameter $\rho\ell$ is temperature independent, with $6.6\times 10^{-16}$ $\Omega$m$^2$ for Cu [2]. Unless otherwise noted, the following parameters are used: RRR (Residual Resistivity Ratio, the ratio of resistivity $\rho$ between room temperature and cryogenic temperature, i.e. 10K) Cu at 50 is chosen. Mean free path of Cu is 40$nm$ at room temperature, and is 2$\mu$m for $RRR=50$ Cu at cryogenic temperature. The film thickness is chosen to be 10$\mu$m. This Cu film at cryogenic temperature is simulated at two frequencies for comparison: 0.1GHz (normal skin effect) and 10.1GHz (anomalous skin effect). For S.S., the room temperature resistivity of $3.6\times 10^{-7}$ $\Omega$m is used.

## IV. Anomalous skin effect of Cu film without substrate

For Cu film without substrate, the surface impedance at 10.1GHz is 6.2+9.9$i$ m$\Omega$ for Cu with $RRR=50$, and is 26.3+27.0$i$ m$\Omega$ for Cu with $RRR=1$. The $E$ field amplitude along the depth of film is shown in Figure 1. With solid line for Cu with $RRR=50$, and dashed line for Cu with $RRR=1$. From this plot one can notice that the $E$ field amplitude decays exponentially along depth for Cu with *RRR*=1, and the surface resistance is close to the surface reactance, it is close to normal skin effect; while for Cu with *RRR*=50, the *E* field amplitude considerably differs from exponential decay [9], and the surface resistance is only 60% of the surface reactance. And more interestingly, at certain depth locations, the *E* field amplitude first decays with depth, and then slightly increases with depth, there appears one or multiple "local minimum". Such "local minimum" also appears in the anomalous skin effect in gas discharged plasmas [14-16].

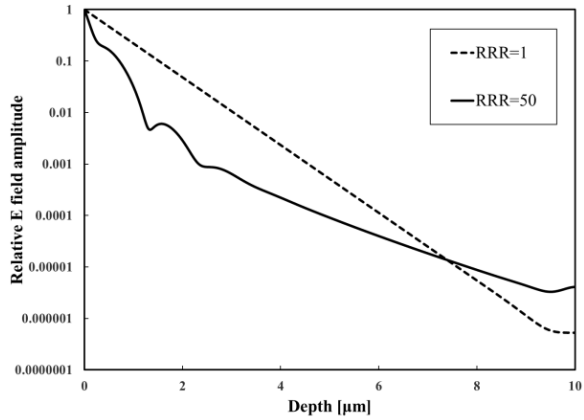

Figure 1. Amplitude of E field along the depth of 10*μm* thick Cu film at 10.1GHz without substrate, with 0 the interface between vacuum and film. Solid line: with 2*μm* mean free path, corresponding to *RRR*=50; Dashed line: with 0.04*μm* mean free path, corresponding to *RRR*=1;

The Re[$EJ^*$] along depth is shown in Figure 2. From the inlet one can clearly see a "local minimum", and Re[$EJ^*$]<0 appears. The physical interpretation is that there are energy exchanges between electromagnetic (EM) field and current carried by electrons, for areas with Re[$EJ^*$]>0, EM field gives energy to current, and for those with Re[$EJ^*$]<0, current gives energy to EM field. The electrons right on the surface can be excited by EM field, and since their mean free path is long, they can penetrate deep into the material, and give energy to EM field. These energy exchanges are represented by the phase relationship between *J(z)* and *E(z)* shown in equation (4). For normal skin effect, *J(z)* is in phase with *E(z)* since $k_a(z)$ is a $\delta$ function, and Re[$EJ^*$] is always larger than 0. The "local minimum", as well as negative Re[$EJ^*$], in the anomalous skin effect, appears to be associated with the field decay that deviates from exponential. Mathematically, if $E=E_0 exp(-z(m+in))$, *m* and *n* are functions of depth *z*, and we consider the simple case that *m'* is a constant negative, here ' is the derivative of *z*, and *m''*=*n'*=*n''*=0, then we have Re[$EJ^*$]=2 $n(m+m'z)EE^*/(\omega\mu_0)$, it can be negative while *z* is big enough.

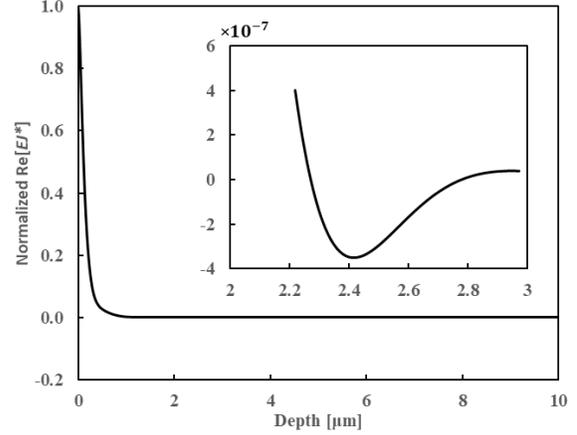

Figure 2. Re[$EJ^*$] along depth for Cu film without substrate at 10.1*GHz*. with inlet plot a zoom-in between 2 and 3*μm*, where "local minimum" and Re[$EJ^*$]<0 appears.

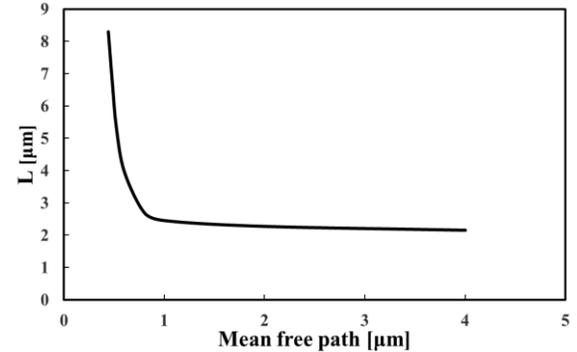

Figure 3. depth *L* of the first point that Re[$EJ^*$]=0 appears versus the mean free path.

To further investigate this phenomenon, the depth *L* of the first point that Re[$EJ^*$]=0 appears is calculated as a function of mean free path, shown in Figure 3 for Cu at 10.1*GHz*. With higher RRR Cu, *L* is closer to the Cu/Vacuum surface. To quantitatively explain the effect of mean free path $\ell$ on the Re[$EJ^*$] for anomalous skin effect, for simplification, three points are evaluated for equation (4) to get the phase relationship between *E* and *J*: point *z*, and two adjacent points *z*±*Δz*, with *Δz* the full width half maximum (FWHM) of $k_a(z)$. $k_a(z)$ is a function of $\ell$, and it is axisymmetrical along *z*=0. We note $k_0=k_a(0)$ and

$k_a(\pm\Delta z) = kk_0 e^{i\varphi}$, and we evaluate the deviation of local current $J(z)$ from its value with the normal skin effect $J_0(z) = k_0 E(z)$:

$$\alpha = \frac{J(z) - J_0(z)}{J_0(z)} = ke^{i\varphi}(e^{\Delta z(m+in)} + e^{-\Delta z(m+in)}) \quad (8)$$

The phase of this coefficient $\alpha$ is $tanh(\Delta zm)tan(\Delta zn)$. With $m$ and $n$ functions of depth $z$, the phase of $\alpha$ oscillates along $z$. And with larger $\ell$, the FWHM $\Delta z$ is larger, this phase oscillates faster, and a smaller $L$ appears. Equation (8) also implies that this deviation oscillates and multiple "local minimums" should appear. In Figure 2, the first "local minimum" appeared at 2.42$\mu m$, shown in the inlet plot, and the second "local minimum", which was not shown in the Figure, appeared at 3.55$\mu m$.

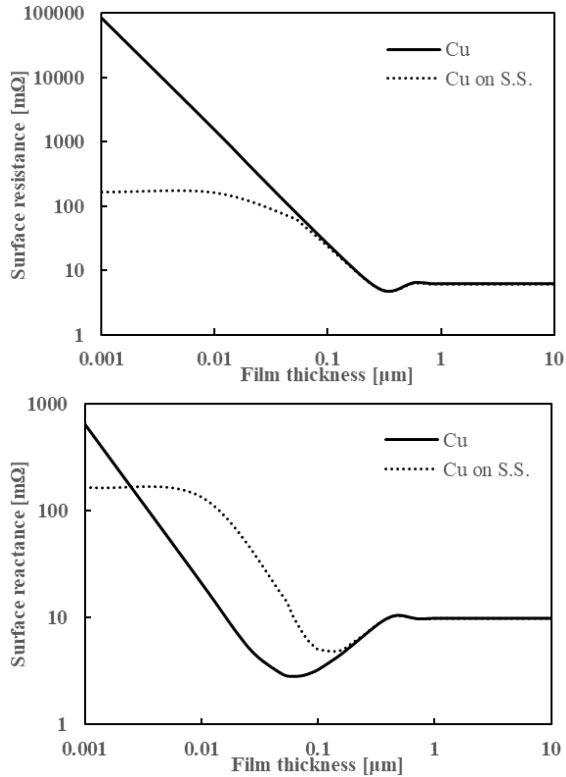

Figure 4. Surface impedance of Cu film versus the film thickness for Cu with 50 RRR at 10.1GHz, with top plot for surface resistance and bottom plot for surface reactance, and solid curves for Cu film without substrate, dotted curves for Cu film on S.S..

For Cu film with the same film thickness, similar to the bulk material, both surface resistance and surface reactance decrease with increasing mean free path $\ell$.

In normal skin effect at dirty limit with $\ell$ close to 0, both surface resistance and surface reactance decrease with increasing film thickness. This is not the case for the anomalous skin effect. For Cu with 50 RRR, the surface impedance versus film thickness is shown in Figure 4 with solid curves, with top plot the surface resistance, and bottom plot the surface reactance. The surface impedance is the same as the surface impedance of bulk material (6.2m$\Omega$ for Cu with the same parameters) while film is thick enough (>1$\mu$m), and while film is thin (<0.1$\mu$m), surface resistance and reactance increase with decreasing film thickness. There appears to be a minimum on surface resistance, with 4.7m$\Omega$ at 0.35$\mu$m thickness. For Cu with 2$\mu$m mean free path at 10.1GHz, if there is no anomalous skin effect, the surface resistance should be 3.6m$\Omega$, while with anomalous skin effect, the mean free path is larger than the effective depth of penetration, causing insufficient sheilding and thus a higher surface resistance at 6.2m$\Omega$. With Cu film thickness smaller than the mean free path, but comparable with the effective depth of penetration, the reflection (difussive in this case) causes an enhancement in sheilding effect, thus a reduction in surface resistance while reducing the film thickness. Combining with the increasing in transmission which causes an increasing in surface resistance, a minimum appears. Please note similar effect was also presented in [6] with a symmetric boundary condition for Aluminum, which is applicable to Cu film without substrate, but not applicable for Cu film with S.S. substrate. For frequncies smaller than 10.1GHz, the effective depth of penetration will be larger, and the film thickness for minimum surface resistance will be larger as well.

## V. Anomalous skin effect of Cu film with S.S. substrate

For Cu film with S.S. substrate, the boundary condition is determined by the surface impedance of S.S.. The $E$ field amplitude of Cu film with S.S. substrate is close to that of Cu film without substrate showing in Figure 1.

The surface impedance versus film thickness is shown in Figure 4 with dotted curves, with top plot the surface resistance, and bottom plot the surface reactance. The surface impedance is the same as the surface impedance of bulk material while film is thick enough (>1$\mu$m), and while film is thin (<0.01$\mu$m), the surface impedance matches that of S.S.. There appears a minimum in surface resistance, with 4.8mΩ at 0.35μm thickness.

As an example, typical parameters for an BNL EIC are shown in Table 1 [17]. The ohmic losses on the pipe walls (in W/m) for a round beam pipe is given by :

$$P = \frac{MQ^2}{2\pi^2 b T_0} \int_0^\infty d\omega R_s(\omega) e^{-(\sigma\omega/c)^2} \quad (9)$$

where $R_s(\omega)$ is the surface resistance of the pipe wall as a function of angular frequency.

For the S.S. beam pipe that is currently using in RHIC, the power loss is 0.56W/m. The power loss versus RRR 50 Cu film thickness is shown in Figure 5. With film thickness thinner than 1μm, the power loss increases with decreasing thickness. Above 4μm thickness, the power loss is the same as the bulk Cu with the same RRR. There is a minimum power loss at 1.2μm thickness, with power loss 3% less than that of bulk. From this plot a conclusion is made that for RRR 50 Cu thin film, the power loss for film thickness more than 1μm should give 3.8% of the power loss for S.S. beam pipe.

Table 1. BNL EIC hadron ring parameters.

| Parameter | Value |
| --- | --- |
| Circumference [m] | 3834 |
| Revolution period $T_0$ [μs] | 12.8 |
| Maximum charge per bunch $Q$ [nC] | 32.8 |
| Minimum rms bunch length $\sigma$ [cm] | 6 |
| Number of bunches in the ring $M$ | 290 |
| Beam pipe radius $b$ [cm] | 3.6 |

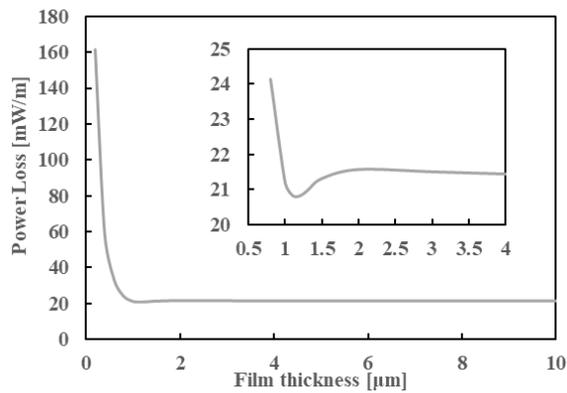

Figure 5. Ohmic power loss per meter versus RRR 50 Cu film thickness, for BNL EIC hadron ring with parameters shown in Table 1, with the inlet on topright corner a zoom-in plot.

## VI. Conclusion

We analyzed the anomalous skin effect for high RRR Cu for high frequency and low temperature application. The amplitude of E field over depth is found to be non-monotonic, which was shown in the previous studies of anomalous skin effect in gas discharged plasmas [14-16]. For Cu film on S.S. substrate, it was found that while film is thin enough, the $R_s$ is close to the value of S.S., and while it is thick enough, it is close to that of bulk Cu. It was also found that while the film thickness is comparable to the effective depth of penetration, a mimimum $R_s$ can be found. This behave can be explained by a combination of enhanced screening effect that causes a reduction in $R_s$, and increasing in transmission that causes an increasing in $R_s$. For BNL EIC application, a minimum of 1μm thin film is required for RRR 50 Cu, with 1.2μm to be the optimimum thickness.

## ACKNOWLEDGEMENT

The work is supported by by Brookhaven Science Associates, LLC under contract No. DE-AC02-98CH10886 with the US DOE. The authors would like to thank J. M. Brennan, S. Verdú-Andrés and A. Zaltsman for useful discussions.